\documentclass[utf8]{frontiersSCNS} 

\usepackage{url,hyperref,lineno,microtype,subcaption}
\usepackage[onehalfspacing]{setspace}

\newcommand{\gl}[1]{\textcolor{black}{#1}}

\def\keyFont{\fontsize{8}{11}\helveticabold }
\def\firstAuthorLast{Lima {et~al.}} 
\def\Authors{Gabriel Lima\,$^{1,2}$, Meeyoung Cha\,$^{2,1,*}$, Chihyung Jeon\,$^{3}$, and Kyung Sin Park\,$^{4}$}
\def\Address{$^{1}$School of Computing, KAIST, Daejeon, Republic of Korea \\
$^{2}$Data Science Group, Institute for Basic Science, Daejeon, Republic of Korea \\
$^{3}$Graduate School of Science and Technology Policy, KAIST, Daejeon, Republic of Korea\\
$^{4}$School of Law, Korea University, Seoul, Republic of Korea}
\def\corrAuthor{55 Expo-ro, Yuseong-gu, Daejeon, Republic of Korea, 34126}

\def\corrEmail{mcha@ibs.re.kr}

\begin{document}
\onecolumn
\firstpage{1}

\title[Conflicts of AI Legal Punishment]{The Conflict Between People's Urge to Punish AI and Legal Systems} 

\author[\firstAuthorLast ]{\Authors} 
\address{} 
\correspondance{} 

\extraAuth{The final version has been published at Frontiers in Robotics and AI. DOI: \url{https://doi.org/10.3389/frobt.2021.756242}}

\maketitle

\begin{abstract}

Regulating artificial intelligence (AI) has become necessary in light of its deployment in high-risk scenarios. This paper explores the proposal to extend legal personhood to AI and robots, which had not yet been examined through the lens of the general public. We present two studies ($N$ = 3,559) to obtain people’s views of electronic legal personhood vis-à-vis existing liability models. Our study reveals people’s desire to punish automated agents even though these entities are not recognized any mental state. Furthermore, people did not believe automated agents’ punishment would fulfill deterrence nor retribution and were unwilling to grant them legal punishment preconditions, namely physical independence and assets. Collectively, these findings suggest a conflict between the desire to punish automated agents and its perceived impracticability. We conclude by discussing how future design and legal decisions may influence how the public reacts to automated agents’ wrongdoings.

\tiny
 \keyFont{ \section{Keywords:} artificial intelligence, robots, AI, legal system, legal personhood, punishment, responsibility} 
\end{abstract}

\section{Introduction}

Artificial intelligence (AI) systems have become ubiquitous in society. To discover where and how these machines\footnote{We use the term ``machine'' as a interchangeable term for AI systems and robots, i.e., embodied forms of AI. Recent work on the human factors of AI systems have used this term to refer to both AI and robots (e.g.,~\citep{kobis2021bad}), and some of the literature that has inspired this research uses similar terms when discussing both entities, e.g.,~\citep{matthias2004responsibility}.} affect people’s lives does not require one to go very far. For instance, these automated agents can assist judges in bail decision-making and choose what information users are exposed to online. They can also help hospitals prioritize those in need of medical assistance and suggest who should be targeted by weapons during war. As these systems become widespread in a range of morally relevant environments, mitigating how their deployment could be harmful to those subjected to them has become more than a necessity. Scholars, corporations, public institutions, and nonprofit organizations have crafted several ethical guidelines to promote the responsible development of the machines affecting the people’s lives~\citep{jobin2019global}. However, are ethical guidelines sufficient to ensure that such principles are followed? Ethics lacks the mechanisms to ensure compliance and can quickly become a tool for escaping regulation~\citep{resseguier2020ai}. Ethics should not be a substitute for enforceable principles, and the path towards safe and responsible deployment of AI seems to cross paths with the law.

The latest attempt to regulate AI has been advanced by the European Union (EU;~\citep{eu2021proposal}), which has focused on creating a series of requirements for high-risk systems (e.g., biometric identification, law enforcement). This set of rules is currently under public and scholarly scrutiny, and experts expect it to be the starting point of effective AI regulation. This research explores one proposal previously advanced by the EU that has received extensive attention from scholars but was yet to be studied through the lens of those most affected by AI systems, i.e., the general public. In this work, we investigate the possibility of extending legal personhood to autonomous AI and robots~\citep{delvaux2017report}.

The proposal to hold machines, partly or entirely, liable for their actions has become controversial among scholars and policymakers. An open letter signed by AI and robotics experts denounced its prospect following the EU proposal (\url{http://www.robotics-openletter.eu/}). Scholars opposed to electronic legal personhood have argued that extending certain legal status to autonomous systems could create human liability shields by protecting humans from deserved liability~\citep{bryson2017and}. Those who argue against legal personhood for AI systems regularly question how they could be punished~\citep{asaro201111,solaiman2017legal}. Machines cannot suffer as punishment~\citep{sparrow2007killer}, nor do they have assets to compensate those harmed.

Scholars that defend electronic legal personhood argue that assigning liability to machines could contribute to the coherence of the legal system. Assigning responsibility to robots and AI could imbue these entities with realistic motivations to ensure they act accordingly~\citep{turner2018robot}. Some highlight that legal personhood has also been extended to other nonhumans, such as corporations, and doing so for autonomous systems may not be as implausible~\citep{van2018we}. As these systems become more autonomous, capable, and socially relevant, embedding autonomous AI into legal practices becomes a necessity~\citep{gordon2020artificial,jowitt2020assessing}. 

We note that AI systems could be granted legal standing regardless of their ability to fulfill duties, e.g., by granting them certain rights for legal and moral protection~\citep{gunkel2018robot,gellers2020rights}. Nevertheless, we highlight that the EU proposal to extend a specific legal status to machines was predicated on holding these systems legally responsible for their actions. Many of the arguments opposed to the proposal also rely on these systems’ incompatibility with legal punishment and pose that these systems should not be granted legal personhood because they cannot be punished. 

\gl{An important distinction in the proposal to extend legal personhood to AI systems and robots is its adoption under criminal and civil law. While civil law aims to make victims whole by compensating them~\citep{prosser1941handbook}, criminal law punishes offenses. Rights and duties come in distinct bundles such that a legal person, for instance, may be required to pay for damages under civil law and yet not be held liable for a criminal offense~\citep{kurki2019theory}. The EU proposal to extend legal personhood to automated systems has focused on the former by defending that they could make ``good any damage they may cause.'' However, scholarly discussion has not been restricted to the civil domain and has also inquired how criminal offenses caused by AI systems could be dealt with~\citep{abbott2020reasonable}.}

\gl{Some of the possible benefits, drawbacks, and challenges of extending legal personhood to autonomous systems are unique to civil and criminal law. Granting legal personhood to AI systems may facilitate compensating those harmed under civil law~\citep{turner2018robot}, while providing general deterrence~\citep{abbott2020reasonable} and psychological satisfaction to victims (e.g., through revenge~\citep{mulligan2017revenge}) if these systems are criminally punished. Extending civil liability to AI systems means these systems should hold assets to compensate those harmed~\citep{bryson2017and}. In contrast, the difficulties of holding automated systems criminally liable extend to other domains, such as how to define an AI system's mind, how to reduce it to a single actor~\citep{gless2016if}, and how to grant them physical independence.}

The proposal to adopt electronic legal personhood addresses the difficult problem of attributing responsibility for AI systems’ actions, i.e., the so-called responsibility gap~\citep{matthias2004responsibility}. Self-learning and autonomous systems challenge epistemic and control requirements for holding actors responsible, raising questions about who should be blamed, punished, or answer for harms caused by AI systems~\citep{de2021four}. The deployment of complex algorithms leads to the ``problem of many things,'' where different technologies, actors, and artifacts come together to complicate the search for a responsible entity~\citep{coeckelbergh2020artificial}. These gaps could be partially bridged if the causally responsible machine is held liable for its actions. 

\gl{Some scholars argue that the notion of a responsibility gap is overblown. For instance,~\citet{johnson2015technology} has asserted that responsibility gaps will only arise if designers choose and argued that they should instead proactively take responsibility for their creations. Similarly, \citet{saetra2021confounding} has argued that even if designers and users may not satisfy all requirements for responsibility attribution, the fact that they chose to deploy systems that they do not understand nor have control over makes them responsible. Other scholars view moral responsibility as a pluralistic and flexible process that can encompass emerging technologies~\citep{tigard2020there}.}

\citet{danaher2016robots} has made a case for a distinct gap posed by the conflict between the human desire for retribution and the absence of appropriate subjects of retributive punishment, i.e., the retribution gap. Humans look for a culpable wrongdoer deserving of punishment upon harm and justify their intuitions with retributive motives~\citep{carlsmith2008psychological}. AI systems are not appropriate subjects of these retributive attitudes as they lack necessary conditions for retributive punishment, e.g., culpability. 

\gl{The retribution gap has been criticized by other scholars, who defend that people could exert control over their retributive intuitions~\citep{kraaijeveld2020debunking} and argue that conflicts between people's intuitions and moral and legal systems are dangerous only if they destabilize such institutions~\citep{saetra2021confounding}. This research directly addresses whether such conflict is real and could pose challenges to AI systems' governance.} Coupled with previous work finding that people blame AI and robots for harm (e.g.,~\citep{kim2006should,malle2015sacrifice,lima2021human,furlough2021attributing,lee2021people}), there seems to exist a clash between people’s reactive attitudes towards harms caused by automated systems and their feasibility. This conflict is yet to be studied empirically.

We investigate this friction. We question whether people would punish AI systems in situations where human agents would typically be held liable. We also inquire whether these reactive attitudes can be grounded on crucial components of legal punishment, i.e., some of its requirements and functions. Previous work on the proposal to extend legal standing to AI systems has been mostly restricted to the normative domain, and research is yet to investigate whether philosophical intuitions concerning the responsibility gap, retribution gap, and electronic legal personhood have similarities with the public view. We approach this research question as a form of experimental philosophy of technology~\citep{kraaijeveld2021experimental}. \gl{This research does not defend that responsibility and retribution gaps are real or can be solved by other scholars' proposals. Instead, we investigate how people's reactive attitudes towards harms caused by automated systems may clash with legal and moral doctrines and whether they warrant attention.}

Recent work has explored how public reactions to automated vehicles (AVs) could help shape future regulation~\citep{awad2018moral}. Scholars posit that psychology research could augment information available to policymakers interested in regulating autonomous machines~\citep{awad2020crowdsourcing}. This body of literature acknowledges that the public view should not be entirely embedded into legal and governance decisions due to harmful and irrational biases. Yet, they defend that obtaining the general public’s attitude towards these topics can help regulators discern policy decisions and prepare for possible conflicts. 

\gl{Viewing the issues of responsibility posed by automated systems as political questions,~\citet{saetra2021confounding} has defended that these questions should be subjected to political deliberation. Deciding how to attribute responsibility comes with inherent trade-offs that one should balance to achieve responsible and beneficial innovation. A crucial stakeholder in this endeavor is those who are subjected to the indirect consequences of widespread deployment of automated systems, i.e., the public~\citep{dewey1927public}. Scholars defend that automated systems ``should be regulated according to the political will of a given community''~\citep{saetra2021research}, where the general public is a major player. Acknowledging the public opinion facilitates the political process to find common ground for successful regulation of these new technologies. If legal responsibility becomes too detached from the folk conception of responsibility, the law might become unfamiliar to those whose behavior it aims to regulate, thus creating the ``law in the books'' instead of the ``law in action''~\citep{brozek2019can}.}

\gl{People's expectations and preconceptions of AI systems and robots have several implications to their adoption, development, and regulation~\citep{cave2019hopes}. For instance, fear and hostility may hinder the adoption of beneficial technology~\citep{cave2018portrayals,bonnefon2020moral}, whereas a more positive take on AI and robots may lead to unreasonable expectations and overtrust---which scholars have warned against~\citep{bansal2019beyond}. Narratives about AI and robots also inform and open new directions for research among developers~\citep{cave2019hopes}, and shape the views of both policymakers and its constituents~\citep{cave2019hopes}.} This research contributes to the maintenance of the ``algorithmic social contract,'' which aims to embed societal values into the governance of new technologies~\citep{rahwan2018society}. By understanding how all stakeholders involved in developing, deploying, and using AI systems react to these new technologies, those responsible for making governance decisions can be better informed of any existing conflicts.


\section{Methods}

\gl{Our research inquired how people's moral judgments of automated systems may clash with existing legal doctrines through a survey-based study. We recruited 3,315 US residents through Amazon Mechanical Turk (see SI for demographic information), who attended a study where they 1) indicated their perception of automated agents' liability and 2) attributed responsibility, punishment, and awareness to a wide range of entities that could be held liable for harms caused by automated systems under existing legal doctrines.}

\gl{We employed a between-subjects study design in which each participant was randomly assigned to a scenario, an agent, and an autonomy level. Scenarios covered two environments where automated agents are currently deployed: medicine and war (see SI for study materials). Each scenario posited three agents: an AI program, a robot (i.e., an embodied form of AI), or a human actor. Although the proposal of extending legal standing to AI systems and robots have similarities, they also have distinct aspects worth noting. For instance, although a ``robot death penalty'' may be a viable option through its destruction, ``killing'' an AI system may not have the same expressive benefits due to varying levels of anthropomorphization. However, extensive literature discuss the two actors in parallel, e.g., ~\citep{turner2018robot,abbott2020reasonable}. We come back to this distinction in our final discussion. Finally, our study introduced each actor as either ``supervised by a human'' or ``completely autonomous.''}

\gl{Participants assigned to an automated agent first evaluated whether punishing it would fulfill some of legal punishment's functions, namely reform, deterrence, and retribution~\citep{solum1991legal,asaro2007robots}. They also indicated whether they would be willing to grant assets and physical independence to automated systems --- two factors that are preconditions for civil and criminal liability, respectively. If automated systems do not hold assets to be taken away as compensation for those they harmed, they cannot be held liable under civil law. Similarly, if an AI system or robot do not possess any level of physical independence, it becomes hard to imagine their criminal punishment. These questions were shown in random order and answered using a 5-point bipolar scale.}

\gl{After answering this set of questions or immediately after consenting to the research terms for those assigned to a human agent, participants were shown the selected vignette in textual format. They were then asked to attribute responsibility, punishment, and awareness to their assigned agent. Responsibility and punishment are closely related to the proposal of adopting electronic legal personhood, while awareness plays a major role in legal judgments (e.g., \textit{mens rea} in criminal law, negligence in civil law). We also identified a series of entities (hereafter associates) that could be held liable under existing legal doctrines, such as an automated system's manufacturer under product liability, and asked participants to attribute the same variables to each of them. All questions were answered using a 4-pt scale. Entities were shown in random order and one at a time.}

\gl{We present the methodology details and study materials in the SI. A replication with a demographically representative sample ($N$ = 244) is also shown in the SI to substantiate all of the findings presented in the main text. This research had been approved by
the first author’s Institutional Review Board (IRB). All data and scripts are available at the project’s repository: \url{https://bit.ly/3AMEJjB}.}


\section{Results}

\begin{figure}
    \centering
    \includegraphics[width=\textwidth]{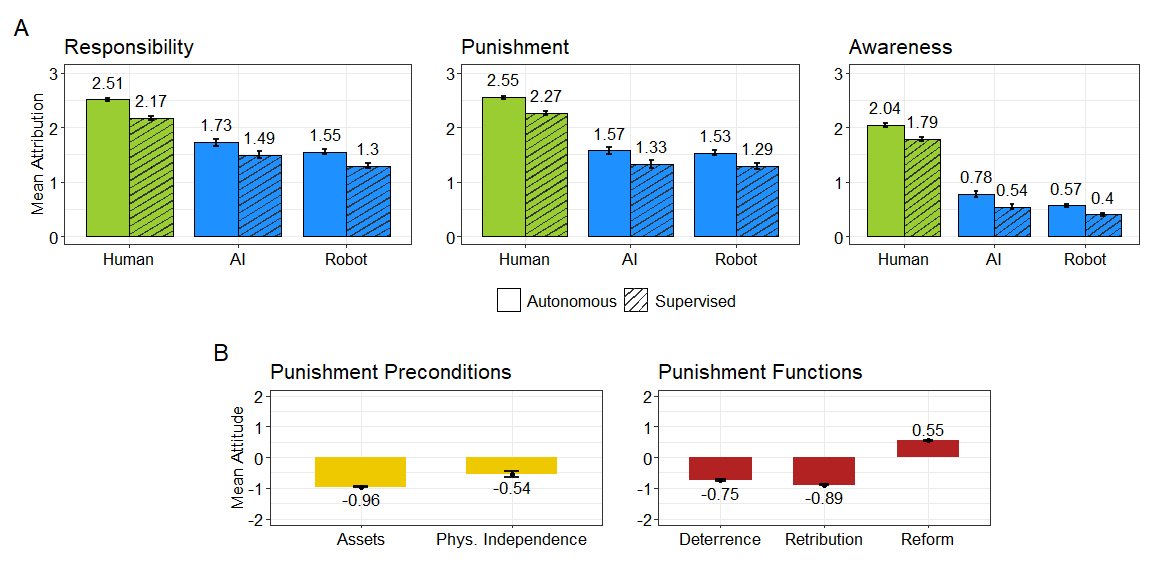}
    \caption{Attribution of responsibility, punishment, and awareness to human agents, AI systems, and robots upon a legal offense (A). Participants’ attitudes towards granting legal punishment preconditions to AI systems and robots (e.g., assets and physical independence) and respondents’ views that automated agents’ punishment would (not) satisfy the deterrence, retributive, and reformative functions of legal punishment (B). Standard errors are shown as error bars.}
    \label{fig:aihuman}
\end{figure}

Figure~\ref{fig:aihuman}A shows the mean values of responsibility and punishment attributed to each agent depending on their autonomy level. Automated agents were deemed moderately responsible for their harmful actions ($M$ = 1.48, $SD$ = 1.16), and participants wished to punish AI and robots to a significant level ($M$ = 1.42, $SD$ = 1.28). In comparison, human agents were held responsible ($M$ = 2.34, $SD$ = 0.83) and punished ($M$ = 2.41, $SD$ = 0.82) to a larger degree. 

A 3 (agent: AI, robot, human) x 2 (autonomy: completely autonomous, supervised) ANOVA on participants’ judgments of responsibility revealed main effects of both agent ($F$(2, 3309) = 906.28, $p <$ .001, $\eta^2_p$  = 0.35) and autonomy level ($F$(1, 3309) = 43.84, $p <$ .001, $\eta^2_p$ =  0.01). The extent to which participants wished to punish agents was also dependent on the agent ($F$(2, 3309) = 391.61, $p <$ .001, $\eta^2_p$ = 0.16) and its autonomy ($F$(1, 3309) = 45.56, $p$ .001, $\eta^2_p$ =  0.01). The interaction between these two factors did not reach significance in any of the models ($p >$ .05). Autonomous agents were overall viewed as more responsible for their actions and deserving of a larger punishment than their supervised counterparts. \gl{We did not observe noteworthy differences between AI systems and robots; the latter were deemed marginally less responsible than AI systems.}

Participants perceived automated agents as only slightly aware of their actions ($M$ = 0.54, $SD$ = 0.88), while human agents were considered somewhat aware ($M$ = 1.92, $SD$ = 1.00). A 3 x 2 ANOVA model revealed main effects for both agent type ($F$(2, 3309) = 772.51, $p <$ .001, $\eta^2_p$ =  0.35) and autonomy level ($F$(1, 3309) = 43.87, $p <$ .001, $\eta^2_p$ =  0.01). The interaction between them was not significant ($p =$ .401). \gl{Robots were deemed marginally less aware of their offenses than AI systems.} Figure~\ref{fig:aihuman}A shows the mean perceived awareness of AI, robots, and human agents upon a legal offense. 
A mediation analysis revealed that perceived awareness of AI systems (coded as -1) and robots (coded as 1) mediated judgments of responsibility (partial mediation, \emph{coef} = -.04, 95\% CI [-.06, -.02]) and punishment (complete mediation, \emph{coef} = -.05, 95\% CI [-.07, .-.02]).  

The leftmost plot of Figure~\ref{fig:aihuman}B shows participants' attitudes towards granting assets and some level of physical independence to AI and robots using a 5-pt scale. These two concepts are crucial preconditions for imposing civil and criminal liability, respectively. Participants were largely contrary to allowing automated agents to hold assets ($M$ = -.96, $SD$ = 1.16) or physical independence ($M$ = -.55, $SD$ = 1.30). Figure~\ref{fig:aihuman}B also shows the extent to which participants believed the punishment of AI and robots might satisfy deterrence, retribution, and reform, i.e., some of legal punishment’s functions. Respondents did not believe punishing an automated agent would fulfill its retributive functions ($M$ = -.89, $SD$ = 1.12) or deter them from future offenses ($M$ = -.75, $SD$ = 1.22); however, AI and robots were viewed as able to learn from their wrongful actions ($M$ = .55, $SD$ = 1.17). We only observed marginal effects ($\eta^2_p \leq$ .01) of agent type and autonomy in participants’ attitudes towards preconditions and functions of legal punishment and present these results in the SI.

The viability and effectiveness of AI systems’ and robots’ punishment depend on fulfilling certain legal punishment’s preconditions and functions. As discussed above, the incompatibility between legal punishment and automated agents is a common argument against the adoption of electronic legal personhood. Collectively, our results suggest a conflict between people’s desire to punish AI and robots and the punishment’s perceived effectiveness and feasibility. 

We also observed that the extent to which participants wished to punish automated agents upon wrongdoing correlated with their attitudes towards granting them assets ($r$(1935) = .11, $p <$ .001) and physical independence ($r$(224) = .21, $p <$ .001). Those who anticipated the punishment of AI and robots to fulfill deterrence ($r$(1711) = .34, $p <$ .001) and retribution ($r$(1711) = .28, $p <$ .001) also tended to punish them more. However, participants’ views concerning automated agents’ reform were not correlated with their punishment judgments ($r$(1711) = -.02, $p$ = .44). In summary, more positive attitudes towards granting assets and physical independence to AI and robots were associated with larger punishment levels. Similarly, participants that perceived automated agents’ punishment as more successful concerning deterrence and retribution also punished them more. Nevertheless, most participants wished to punish automated agents regardless of the punishment’s infeasibility and unfulfillment of retribution and deterrence.

\begin{figure}
    \centering
    \includegraphics[width=\textwidth]{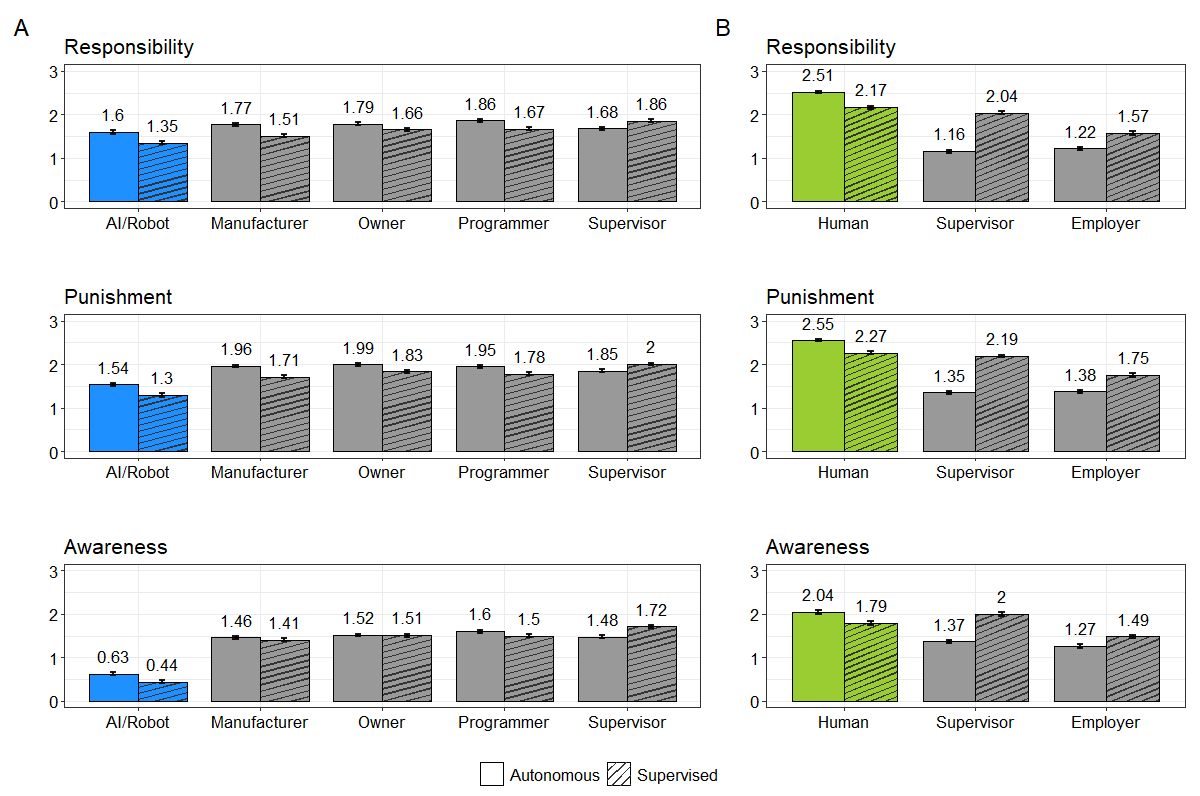}
    \caption{Attribution of responsibility, punishment, and awareness to AI systems, robots, human agents, and entities that could be held liable under existing doctrines (i.e., associates; A). Assignment of responsibility, punishment, and awareness to human agents and corresponding associates (B). Standard errors are shown as error bars.}
    \label{fig:associates}
\end{figure}

Participants also judged a series of entities that could be held liable under existing liability models concerning their responsibility, punishment, and awareness for an agent’s wrongful action. All of the automated agents’ associates were judged responsible, deserving of punishment, and aware of the agents’ actions to a similar degree (see Figure~\ref{fig:associates}). The supervisor of a supervised AI or robot was judged more responsible, aware, and deserving of punishment than that of a completely autonomous system. In contrast, attributions of these three variables to all other associates were larger in the case of an autonomous agent. In the case of human agents, their employers and supervisors were deemed more responsible, aware, and deserving of punishment when the actor was supervised. We observed the opposite effect for the human agents. We present a complete statistical analysis of these results in the SI. 

\section{Discussion}

Our findings demonstrate a conflict between participants’ desire to punish automated agents for legal offenses and their perception that such punishment would not be successful in achieving deterrence or retribution. This clash is aggravated by participants’ unwillingness to grant AI and robots what is needed to legally punish them, i.e., assets for civil liability and physical independence for criminal liability. This contradiction in people’s moral judgments suggests that people wish to punish AI and robots even though they believe that doing so would not be successful, nor are they willing to make it legally viable. 

These results are in agreement with Danaher’s (\citeyear{danaher2016robots}) retribution gap. Danaher acknowledges that people might blame and punish AI and robots for wrongful behavior due to humans’ retributive nature, although they may be wrong in doing so. Our data implies that Danaher’s concerns about the retribution gap are significant and can be extended to other considerations, i.e., deterrence and the preconditions for legal punishment. Past research shows that people also ground their punishment judgments in functions other than retribution~\citep{twardawski2020all}. Public intuitions concerning the punishment of automated agents are even more contradictory than previously advanced by Danaher: they wish to punish AI and robots for harms even though their punishment would not be successful in achieving some of legal punishments' functions or even viable, given that people would not be willing to grant them what is necessary to punish them.

\gl{Our results show that even if responsibility and retribution gaps can be easily bridged as suggested by some scholars~\citep{saetra2021confounding,tigard2020there,johnson2015technology}, there still exists a conflict between the public reaction to harms caused by automated systems and their moral and legal feasibility. 
The public is an important stakeholder in the political deliberation necessary for beneficial regulation of AI and robots, and their perspective should not be rejected without consideration. An empirical question that our results pose is whether this conflict warrants attention from scholars and policymakers, i.e., if they destabilize political and legal institutions~\citep{saetra2021confounding} or leads to lack of trust in legal systems~\citep{abbott2020reasonable}. For instance, it may well be that the public may need to be taught to exert control over their moral intuitions, as suggested by~\citet{kraaijeveld2020debunking}.}

Although participants did not believe punishing an automated agent would satisfy the retributive and deterrence aspects of punishment, they viewed robots and AI systems as capable of learning from their mistakes. Reform may be the crucial component of people’s desire to punish automated agents. Although the current research might not be able to clear this inquiry, we highlight that future work should explore how participants imagine the reform of automated agents. Reprogramming an AI system or robots can prevent future offenses, yet it will not satisfy other indirect reformative functions of punishment, e.g., teaching others that a specific action is wrong. Legal punishment, as it stands, does not achieve the reprogramming necessary for AI and robots. Future studies may question how people’s preconceptions of automated agents’ reprogramming influence people’s moral judgments. 

It might be argued that our results are caused by how the study was constructed. For instance, participants who attributed large punishment levels to automated agents might have reported being more optimistic about its feasibility so that their responses become compatible. However, we observe trends that methodological biases cannot explain but can only result from participants’ a priori contradiction (see SI for detailed methodology). This work does not posit this contradiction as a universal phenomenon; we observed a significant number of participants attributing no punishment whatsoever to electronic agents. Nonetheless, we observed similar results in a demographically representative sample of respondents (see SI).

\gl{We did not observe significant differences in responses to how much AI systems and robots should be punished. The differences in responsibility and awareness judgments were marginal and likely affected by our large sample size. As discussed above, there are different challenges when adopting electronic legal personhood for AI and robots. Embodied machines may be easier to punish criminally if legal systems choose to do so, for instance through the adoption of a ``robot death penalty.'' Nevertheless, our results suggest that the conflict between people's moral intuitions and legal systems may be independent of agent type. Our study design did not control for how people imagined automated systems, which could have affected how people make moral judgments about machines. For instance, previous work has found that people evaluate the moral choices of a human-looking robot as less moral than humans' and non-human robots' decisions~\citep{laakasuo2021moral}.}

Regardless of respondents’ judgments of responsibility and punishment concerning AI and robots, people largely viewed them as unaware of their actions. Much human-computer interaction research has focused on developing social robots that can elicit mind perception through anthropomorphization~\citep{waytz2014mind,darling2016extending}. Therefore, we may have obtained higher perceived awareness had we introduced what the robot or AI looked like, which in turn could have affected respondents’ responsibility and punishment judgments, as suggested by~\citet{bigman2019holding} and our mediation analysis. \gl{These results may also vary by the actor, as robots are subject to higher levels of anthropomorphization.} Past research has also shown that if an AI system is described as an anthropomorphized agent rather than a mere tool, it is attributed more responsibility for creating a painting~\citep{epstein2020gets}. A similar trend was observed with autonomous AI and robots, who were assigned more responsibility and punishment than supervised agents, as previously found in the case of autonomous vehicles~\citep{awad2020drivers} and other scenarios~\citep{furlough2021attributing,kim2006should}.

\subsection{The Importance of Design, Social, and Legal Decisions}

The respondents’ attitudes concerning the fulfillment of punishment preconditions and functions by automated agents were correlated with the extent to which participants wished to punish AI and robots. This finding suggests that people’s moral judgments of automated agents’ actions can be nudged based on how their feasibility is introduced. 

For instance, to clarify that punishing AI and robots will not satisfy human needs for retribution, will not deter future offenses, or is unviable given they cannot be punished similarly to other legal persons may lead people to denounce automated agents’ punishment. If legal and social institutions choose to embrace these systems, e.g., by granting them certain legal status, nudges towards granting them certain perceived independence or private property may affect people’s decision to punish them. Future work should delve deeper into the causal relationship between attitudes towards the topic and people’s attribution of punishment to automated agents. 

Our results highlight the importance of design, social, and legal decisions in how the general public may react to automated agents. Designers should be aware that developing systems that are perceived as aware by those interacting with them may lead to heightened moral judgments. For instance, the benefits of automated agents may be nullified if their adoption is impaired by unfulfilled perceptions that these systems should be punished. Legal decisions concerning the regulation of AI and their legal standing may also influence how people react to harms caused by automated agents. Social decisions concerning how to insert AI and robots into society, e.g., as legal persons, should also affect how we judge their actions. People’s preconceptions of these systems’ capabilities and roles are crucial components of the public’s reactions to their wrongdoing. Future decisions should be made carefully to ensure that laypeople's reactions to harms caused by automated systems do not clash with regulatory efforts.

\section{Concluding Remarks}

Electronic legal personhood grounded on automated agents’ abilities to fulfill duties does not seem a viable path towards the regulation of AI. This approach can only become an option if AI and robots are granted assets or physical independence, which would allow civil or criminal liability to be imposed, or if punishment functions and methods are adapted to AI and robots. People’s intuitions about automated agents’ punishment are somewhat similar to scholars who oppose the proposal. However, a significant number of people still wish to punish AI and robots independently of their a priori intuitions.

By no means this research proposes that robots and AI should be the sole entities to hold liability for their actions. In contrast, responsibility, awareness, and punishment were assigned to all associates. We thus posit that a distributed liability assignment among all entities involved in deploying these systems would follow the public perception of the issue. Such a model could take joint and several liability models as a starting point by enforcing the proposal that various entities should be held jointly liable for damages.

Our work also raises the question of whether people wish to punish AI and robots for reasons other than retribution, deterrence, and reform. For instance, the public may punish electronic agents not for their deterrence but general or indirect deterrence~\citep{twardawski2020all}. Punishing an AI could educate humans that a specific action is wrong without the negative consequences of human punishment. Recent literature in moral psychology also proposes that humans might strive for a morally coherent world, where seemingly contradictory judgment patterns arise so that public perception of agents’ moral qualities match the moral qualities of their actions’ outcomes~\citep{clark2015moral}. We highlight that legal punishment is not only directed at the wrongdoer but also fulfills other functions in society that future work should inquire about when dealing with automated agents. Finally, our work poses the question of whether proactive actions towards holding existing legal persons liable for harms caused by automated agents would compensate for people's desire to punish them. For instance, future work might examine whether punishing a system's manufacturer may decrease the extent to which people punish AI and robots. \gl{Even if the responsibility gap can be easily solved, conflicts between the public and legal institutions might continue to pose challenges to the successful governance of these new technologies.}

We selected scenarios from active areas of AI and robotics (i.e., medicine and war; see SI). The moral judgment of electronic agents’ actions might change depending on the scenario or background. The proposed scenarios did not introduce, for the sake of feasibility and brevity, much of the background usually considered when legally judging someone’s actions. We did not control for any previous attitudes towards AI and robots or knowledge of related areas, such as law and computer science, which could result in different judgments among the participants. 

This research has found a contradiction in people’s moral judgments of AI and robots: they wish to punish automated agents, although they know that doing so is not legally viable nor successful. We do not defend the thesis that automated agents should be punished for legal offenses or have their legal standing recognized. Instead, we highlight that the public's preconceptions of AI and robots influence how they react to their harmful consequences. Most crucially, we showed that people’s reactions to these systems’ failures might conflict with existing legal and moral systems. Our research showcases the importance of understanding the public opinion concerning the regulation of AI and robots. Those making regulatory decisions should be aware of how the general public may be influenced or clash with such commitments.

\section*{Conflict of Interest Statement}

The authors declare that the research was conducted in the absence of any commercial or financial relationships that could be construed as a potential conflict of interest.

\section*{Author Contributions}

All authors designed the research. GL conducted the research. GL analyzed the data. GL wrote the paper, with edits from MC, CJ, and KS.

\section*{Funding}
This research was supported by the Institute for Basic Science (IBS-R029-C2).



\section*{Data Availability Statement}
The datasets generated for this study and the scripts for analysis are available at (\url{https://bit.ly/3AMEJjB}).

\bibliographystyle{frontiersinSCNS_ENG_HUMS} 
\bibliography{test}

\end{document}